\documentclass[%
 aip, rsi, reprint, amsmath, amssymb]{revtex4-2}

\usepackage[separate-uncertainty=true, multi-part-units=single]{siunitx} 
\usepackage{braket}		
\usepackage{graphicx}
\usepackage{dcolumn}
\usepackage{bm}
\usepackage{multirow}
\usepackage[utf8]{inputenc}
\usepackage[T1]{fontenc}
\usepackage{mathptmx}
\usepackage{etoolbox}
\usepackage{bbold}
\usepackage{xcolor}

\makeatletter
\def\@email#1#2{%
 \endgroup
 \patchcmd{\titleblock@produce}
  {\frontmatter@RRAPformat}
  {\frontmatter@RRAPformat{\produce@RRAP{*#1\href{mailto:#2}{#2}}}\frontmatter@RRAPformat}
  {}{}
}%
\makeatother
\begin{document}

\title[Robotized polarization characterization platform for free-space quantum communication optics]{Robotized polarization characterization platform for free-space quantum communication optics}
\author{Youn Seok Lee}
\affiliation{Institute for Quantum Computing and Department of Physics and Astronomy, University of Waterloo, Waterloo, Ontario N2L 3G1, Canada}
  \email{ys25lee@uwaterloo.ca}
  
\author{Kimia Mohammadi}
 \affiliation{Institute for Quantum Computing and Department of Physics and Astronomy, University of Waterloo, Waterloo, Ontario N2L 3G1, Canada}

\author{Lindsay Babcock}
\affiliation{Department of Physics and Astronomy, Simon Fraser University, Burnaby, British Columbia V5A 1S6, Canada}

\author{Brendon L. Higgins}
\affiliation{Institute for Quantum Computing and Department of Physics and Astronomy, University of Waterloo, Waterloo, Ontario N2L 3G1, Canada}

\author{Hugh Podmore}
\affiliation{Honeywell Aerospace, 303 Terry Fox Dr, Ottawa, Ontario K2K 3J1, Canada}

\author{Thomas Jennewein}
\affiliation{Institute for Quantum Computing and Department of Physics and Astronomy, University of Waterloo, Waterloo, Ontario N2L 3G1, Canada}

\date{\today}

\begin{abstract}
We develop a polarization characterization platform for optical devices in free-space quantum communications. We demonstrate an imaging polarimeter, which analyzes both incident polarization states and the angle of incidence, attached to a six-axis collaborative robot arm, enabling polarization characterization at any position and direction with consistent precision. We present a detailed description of each subsystem including the calibration and polarization-test procedure, and analyze polarization-measurement errors caused by imperfect orientations of the robot arm using a Mueller-matrix model of polarimeters at tilt incidence. We perform a proof-of-principle experiment for an angle-dependent polarization test for a commercial silver-coated mirror for which the polarization states of the reflected light can be accurately calculated. Quantitative agreement between the theory and experiment validates our methodology. We demonstrate the polarization test for a \SI{20.3}{\centi\meter} lens designed for a quantum optical transmitter in Canada's Quantum Encryption and Science Satellite (QEYSSat) mission.
\end{abstract}

\maketitle

\section{\label{sec:intro}Introduction}

Over the past decade, a number of experiments have demonstrated quantum communications to various moving platforms, such as hot-air balloon~\cite{Wang2013}, truck~\cite{Truck2015}, aircraft~\cite{Pugh_2017, Nauerth2013}, and drone~\cite{Drone2020}. In particular, with achievements in China’s Quantum Experiments at Space Scale mission~\cite{Liao2017,Yin2017_eps_qkd,Yin2020}, quantum communications using satellites provide a platform for global-scale quantum key distributions as well as fundamental quantum optics experiments in a relativistic length scale. Several countries are endeavouring to create quantum links between ground and space in various scenarios~\cite{Bedington2017,sidhu2021advances}. In Canada, the Quantum Encryption and Science Satellite (QEYSSat) mission has been developing a satellite payload and ground stations with the objectives of long-distance quantum key distributions (QKD) and long-distance quantum entanglement tests via the exchange of polarized photons in an uplink configuration~\cite{Jennewein_qeyssat2018,Podmore2019,Podmore2021}.

\begin{figure*}[t]
	\centering
	\includegraphics[width=0.8\linewidth]{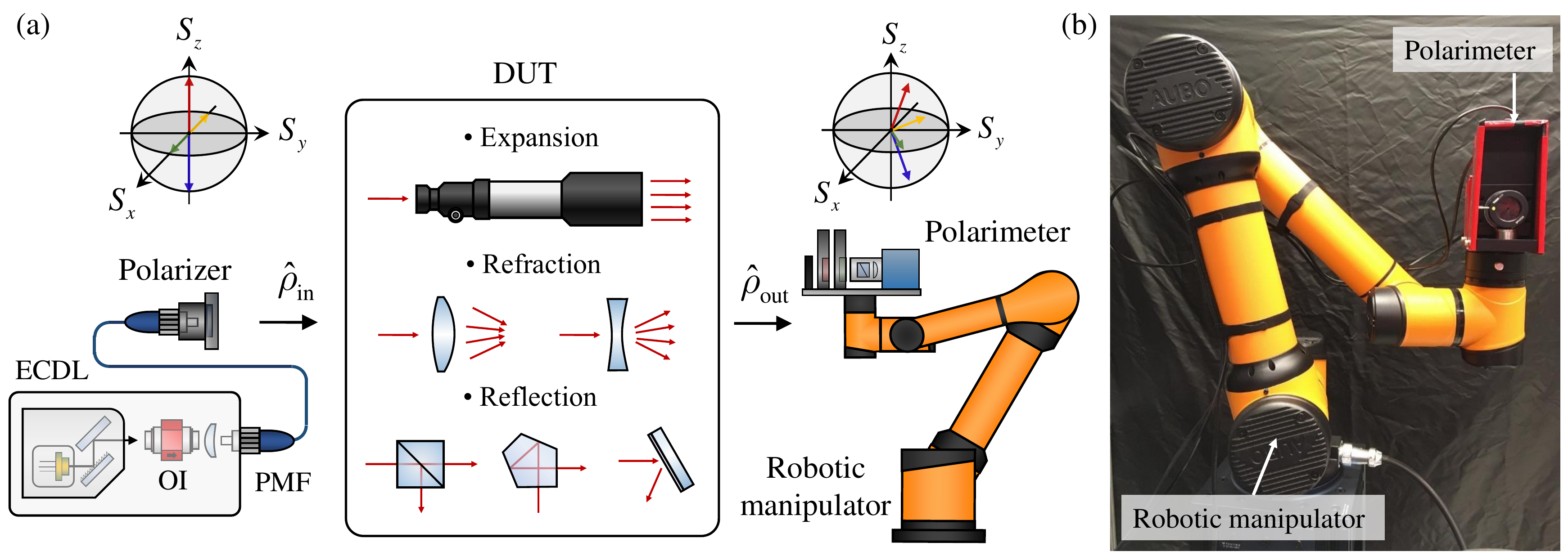}
	\caption{(a) Conceptual design of the polarization characterization system; ECDL, external cavity diode laser; OI, optical isolator; PMF, polarization-maintaining fiber. Four linearly polarized states $\hat{\rho}_{\text{in}}$ are injected into the device under test (DUT) and the polarization states of the transmitted or reflected light $\hat{\rho}_{\text{out}}$ is examined. The output beam size and direction varies depending on the type of optics. (b) Photographs of the robotized polarization-characterization system.
	}
	\label{fig:conceptual_description}
\end{figure*}

Encoding quantum information in optical polarization is a straightforward and robust approach to free-space quantum communication, but depends critically on preservation of high-purity polarized states of light throughout the optical chain. Degradation of polarization quality necessarily impacts the performance and any protocol, such as QKD, being attempted. In particular, free-space communications with moving platforms require specialized photon transceivers to create efficient quantum channels. The transceivers typically consist of a large ``front-end" telescope (pointed at the other telescope) supported by small ``back-end" optics for multiplexing of quantum light with a strong beacon, fine-pointing actuation, etc.~\cite{Wang2013,Pugh_2017,Podmore2019,Podmore2021}. The preservation of polarization states throughout such complex optical terminals is challenging because some polarization rotations or depolarizations are fundamentally inevitable and are easily caused by myriad reasons, e.g., stress-induced birefringence, or thermal expansion of birefringent material. Also, optical coatings usually cause phase shifts of polarizations, which could lead depolarizations when the phase shifts are not uniformly applied across the spatial profile of the incident beam. Moreover, the polarization effect depends on the geometry of the optical terminal; mere reflection/refraction changes the polarization state depending on the incident angle. Therefore, great care must be taken to design optical terminals to preserve the polarization state and it is essential to ensure polarization is preserved at the major interfaces, and the effect of any individual component on polarization is both well understood and verified.

Characterization of the polarization effect of an optical element is performed by injecting known polarization states and measuring the outcomes. The polarization testbed must be capable of precise generation and accurate measurement of polarization states. Especially for devices in free-space quantum communications, the task must be performed for the optical elements of small to large sizes which constitute the quantum optical terminals. Moreover, the polarization state analysis must be attained at the range of orientations and positions over which the terminal's components will operate. This is challenging with commercially available polarimeters because they are typically optimized for a small field of view (FoV) with a limited beam aperture, necessitating significant modification of the testbed for each test optic. As a consequence, most prior works limited their focus, such as on telescopes~\cite{Wu2017} or for an entire assembly in an end-to-end manner~\cite{Han2020,Wu2020}.

Various methods have been developed to characterize polarization effects. For relatively small optics, the angle-dependent polarization test can be achieved by ellipsometry-like methods~\cite{Sankarasubramanian1999,Sun2014,Gu2018}. However, most existing methods are specialized for non-divergent elements. Large telescopes or mirrors often require a specialized test platform. One conventional method for polarization tests on a large telescope, developed for solar observatories, is to build a similar-sized calibration unit in front of the aperture consisting of arrays of rectangular foils which transmit linearly polarized light using sunlight as input~\cite{Almeida1991,Kiyohara2004,Beck2005,Ichimoto2008}. Although fast and simple for outdoor telescopes, such an approach requires large calibrated optics and is not suitable for component testing or indoor operation. Futhermore, the approach is designated only for telescopes—the test setup is not adaptive to other large optical elements such as lenses and curved mirrors. 

Here, we develop a polarization characterization platform for optical devices in free-space quantum communications which can be used indoors and accommodate a wide range of front- and back-end optics. The system utilizes a six-axis collaborative robot arm that moves a polarimeter to analyze the polarization state of light at desired positions and angles. The robot arm was identified as the most cost effective solution, as it enables precise motion over a \SI{0.9}{\meter} range, with the ability to control the measurement device in all six degrees of freedom. To show the suitability of our approach we performed a detailed error analysis on the motional precision of the robot arm, and furthermore designed the polarimeter monitor the angle of incidence (AOI) during the polarization test. The system makes it possible to test inch-sized optics to half-meter-diameter optics (or larger) at consistent accuracy and exhibits great repeatability. The characterization process is fully automated, including the robot's trajectory, data collection and analysis.

This paper is organized as follows. In section~\ref{sec:methodology}, we describe our polarization-test setup and the conceptual design for our polarimeter. In section~\ref{sec:design_analysis}, we provide a detailed analysis of the polarimeter with Mueller matrices as well as our calibration method for the polarimeter. In section~\ref{sec:setup}, we present our experimental setup and the coordinate alignment procedures for the robot arm. In section~\ref{sec:results}, we perform a proof-of-principle experiment of the polarization characterization for a commercially available silver-coated mirror and a \SI{20.3}{\centi\meter} lens custom-designed to support the QEYSSat mission. The angle-dependent polarization effect of the mirror is accurately modelled by multilayer thin-film coating calculations, and comparison of the experimental results with the theoretical predictions validate the measurement system. Concluding remarks are given in section 6.

\section{\label{sec:methodology}Methodology}

\begin{figure*}[t]
	\centering
	\includegraphics[width=0.8\linewidth]{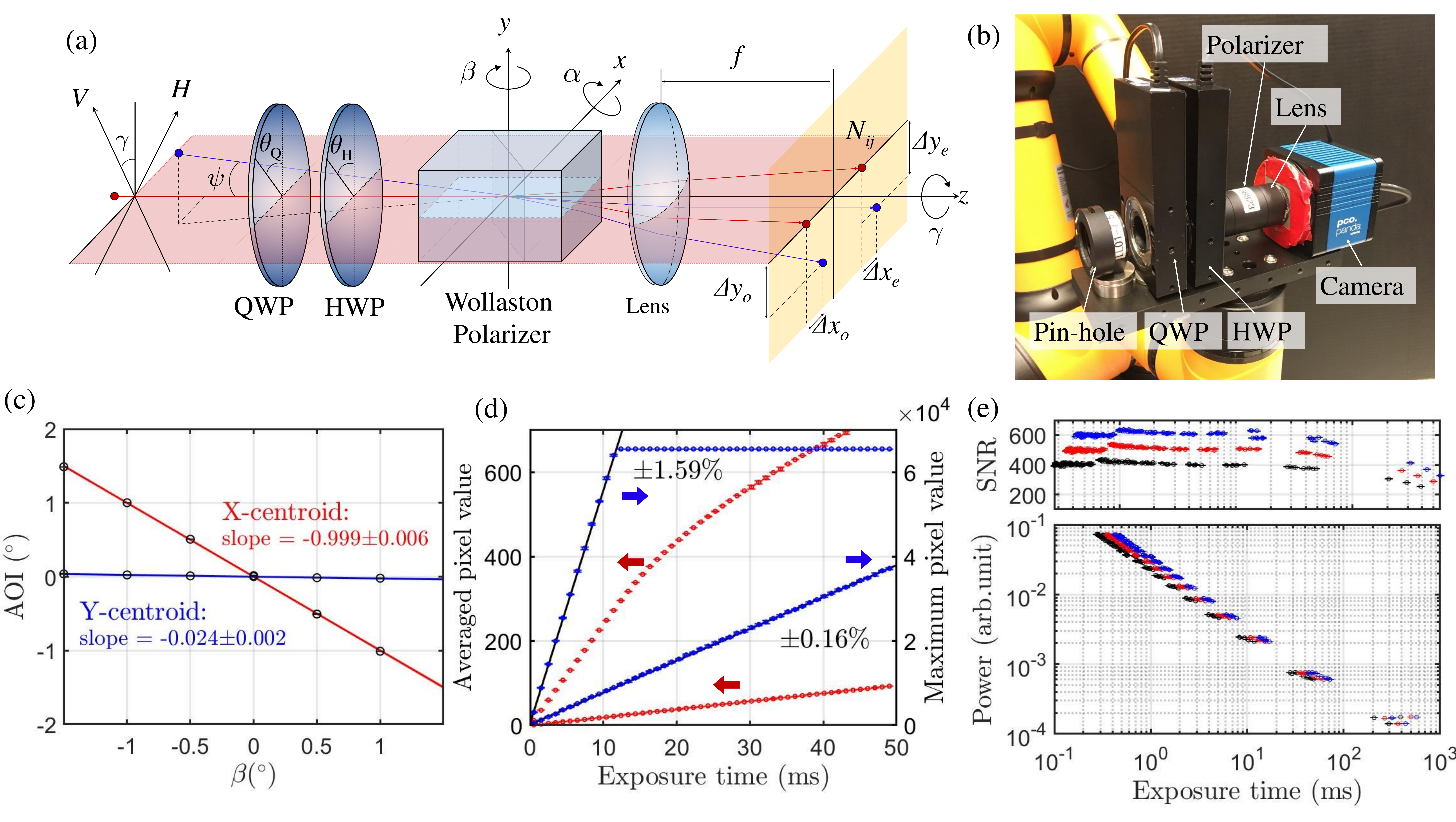}
	\caption{(a) A schematic diagram for the optical configuration of our polarimeter; HWP, half-wave plate; QWP, quarter-wave plate. (b) A close up of the polarimeter attached to the end effector of the robotic manipulator. Experimental characterization of the polarimeter in terms of (c) the angle of incidence via centroid measurements, (d) the linearity of the camera exposure time, and (e) the dynamic range of the optical power measurement. The red and blue arrows point to the corresponding vertical scale for the maximum and averaged pixel values.}
	\label{fig:polarimeter_description}
\end{figure*}

Figure~\ref{fig:conceptual_description} illustrates the concept of our polarization characterization platform. We inject several well-deﬁned polarization states into a device under test (DUT) and perform polarization state tomography on the output states to see how they change. To match our intended application, we consider four incident polarization states: $\ket{H}$ (horizontal), $\ket{V}$ (vertical), $\ket{D}$ (diagonal, \SI{45}{\degree}), and $\ket{A}$ (antidiagonal, \SI{-45}{\degree}). This is sufficient to determine performance of the system in the context of BB84-style QKD protocol\footnote{To perform full process tomography, such as to establish Mueller matrices, one could straightforwardly incorporate additional circularly polarized incident states.}. In our system, the light source is an external cavity diode laser (DLpro, Toptica photonics) operating in continuous-wave mode at \SI{785}{\nano\meter} wavelength, which produces a stable intensity for the polarization test. The input polarization state is initially determined by an optical isolator and delivered through a polarization-maintaining fiber. Upon exiting the fiber, the laser light passes through a linear-film polarizer (LPVIS100, Thorlabs) with its polarization axis aligned to the polarization of the incident field to maximize the transmitted power. The fiber and the polarizer are mounted on a motorized rotation stage (PR50PP, Newport) which rotates both elements altogether to define each of the four input polarization states in turn with accuracy of $\pm$\SI{0.025}{\degree}. Once the light exits the test optic, it reaches a polarimeter which is attached to a six-joint robotic manipulator (AUBO-i5, AUBO Robotics). The robotic manipulator moves the polarimeter to measure the output polarization state at the desired positions and angles. 

Our robotic manipulator is driven by six geared servo-motors with absolute encoders at all joints. This robot has a reach of \SI{0.924}{\meter}, which can easily scan over the entire trajectory around the optics being tested. According to the manufacturer's speciﬁcations, the robot has position repeatability of \SI{0.02}{\milli\meter} and position accuracy of \SI{2}{\milli\meter}. The average orientation repeatability and accuracy are \SI{0.004}{\degree} and \SI{0.5}{\degree}, respectively. A pre-programmed teaching pendant supports manual control by the touch-screen tablet interface, and the C++ SDK allows us to remotely control the robot arm.

Figure~\ref{fig:polarimeter_description}(a) shows the schematic diagram of the polarimeter. It consists of a quarter- and half-wave plate (QWP and HWP), a Wollaston polarizer (68-823, Edmund Optics), a convex lens ($f=$\SI{30}{\milli\meter}, 49-115, Edmund Optics), and a CMOS camera (pco.panda.4.2, PCO). The two waveplates are mounted on motorized rotation stages (PR50PP, Newport) while the principal axis of the Wollaston polarizer is fixed in our polarimeter. The two waveplates and polarizer project the incident polarization state onto six tomographically complete polarization states ($\ket{H}$, $\ket{V}$, $\ket{D}$, $\ket{A}$, $\ket{R}$ right-circular, and $\ket{L}$ left-circular). The projected states are converted to the intensity distributions at the output port of the Wollaston polarizer. A convex lens is placed after the polarizer such that the far-field image of the intensity distribution is mapped on the focal plane where the CMOS imaging sensor is placed. The size of our imaging sensor is $13.3 \times 13.3$\si{\milli\meter\squared} and includes $2048\times2048$ pixels of size $6.5 \times 6.5$\si{\micro\meter\squared}. In this way, the variation of the AOI appears as the translation of the intensity distribution by $\Delta x_{e(o)}$ and $\Delta y_{e(o)}$ which can be precisely measured from image pixel values $N$ by sub-pixel centroid algorithms~\cite{subpixelAccuracy}.

We consider the variation of the AOI that mainly comes from the imperfect orientation of the robotic end effector which is defined by three rotation angles ($\alpha$,$\beta$,$\gamma$) around the three orthogonal axes ($x$,$y$,$z$), as shown in Figure~\ref{fig:polarimeter_description}(a). The angular deviation under consideration is on the order of \SI{0.5}{\degree} and aberrations of the imaging lens are neglected. As the split angle between the ordinary and extraordinary light exiting the polarizer depends on its tilt angle~\cite{Simon:86}, we calculate the central point of the two centroids $\Delta x = (\Delta x_{e} + \Delta x_{o})/2$ and $\Delta y = (\Delta y_{e} + \Delta y_{o})/2$ to cancel such effect. Thus, under the paraxial approximation for the lens, the centroid shifts $\Delta x$ and $\Delta y$ are related to the AOI by the formula
\begin{eqnarray}\label{eq:centroid}
\centering
\Delta x &= f\tan\beta\\
\Delta y &=f\tan\alpha.
\end{eqnarray}
Note that the $\gamma$ rotation is not directly detected by measuring the shift of the intensity distribution as it rather appears as the variation of the intensity values because the polarizer and the camera are rotated altogether.

Although the usage of the camera provides accurate estimation of the AOI, one major drawback of such an imaging polarimeter is the limited dynamic range of optical power measurements with the camera. The issue becomes significant especially when the polarization measurement basis is aligned to the incident polarization axis. For example, our camera exhibits dynamic range of 21,500:1, an order of magnitude smaller than the extinction ratio of the Wollaston polarizer (200,000:1). The signal-to-noise ratio (SNR) of the captured images directly impacts the precision of measuring polarization states. For a given camera with its quantum efficiency $\eta$ and the exposure time $\Delta T$, the SNR is given by
\begin{equation}
\text{SNR} = \frac{P\eta\Delta T}{\sqrt{N_{\text{shot}}^{2}+I_{\text{dark}}\Delta T+N_{\text{readout}}^{2}}},
\end{equation}
where $P$ is the optical power of the incident light, $N_{\text{shot}}=\sqrt{P\eta\Delta T}$ is the shot noise, $I_{\text{dark}}$ is the dark current, and $N_{\text{readout}}$ is the readout noise. For sufficient optical incident power and long exposure time $\Delta T$, the SNR is mainly determined by shot noise. For a given waveplate setting, we capture two images and adjust the exposure time before each image acquisition to measure the optical power of ordinary and extraordinary fields separately with acceptable SNR. The optical power detected by a pixel at $i$-th row and $j$-th column of the imaging sensor is obtained by the measured pixel value $N_{ij}$ divided by the exposure time $\Delta T$. The total incident power $P_{\text{meas}}$ is then estimated by averaging the power value over a region surrounding the focal spot of the incident field as
\begin{equation} 
P_{\text{meas}} = \frac{1}{\Delta T}\bigg[\sum_{i,j=1}^{n}\frac{N_{ij}}{n} - \sum_{i,j=1}^{m}\frac{N_{ij}}{m}\bigg].
\end{equation}
Here, we subtracted background noise to calculate the net power values. The background noise is estimated by the averaged power over the outside of the bright region. $n$ and $m$ are the number of pixels used to estimate incident power values and background noises, respectively. We determine the measured polarization states by evaluating for each Stokes vector $\vec{S} = [ S_{0} , S_{1} , S_{2} , S_{3} ]^\top$, where $S_{0}$ is the total power of the incident light, $S_{1}$ denotes the bias for $\ket{H}$ and $\ket{V}$, $S_{2}$ for $\ket{D}$ and $\ket{A}$, and $S_{3}$ for $\ket{R}$ and $\ket{L}$.

We characterized our polarimeter in terms of the accuracy of the centroid and power measurement as well as the dynamic range. First, we mounted the polarimeter on the robotic manipulator, injected laser light in a fixed propagation direction, and measured the centroids $\Delta x$ and $\Delta y$ as a function of the rotation $\beta$ which are then converted to the AOI via Eq.~\ref{eq:centroid}. The slope is estimated to be nearly unity in $x$--axis via least-square fitting with \SI{0.6}{\percent} standard error of the regression, indicating accurate AOI measurement, as shown in Figure~\ref{fig:polarimeter_description}(c). Secondly, with a constant incident optical power, we recorded the maximum pixel values as a function of the camera exposure time to ensure linearity of the exposure time control, as shown in Figure~\ref{fig:polarimeter_description}(d). The slope is estimated by the same fitting method, and the relative uncertainty of the power measurement is estimated to be around \SI{1}{\percent}. Finally, the dynamic range is characterized by varying the incident power. We varied the incident optical power while allowing automated control of the camera exposure time to maintain a constant SNR over the range of incident power, as shown in Figure~\ref{fig:polarimeter_description}(e). The optical power was measured over a range of three orders of magnitude while maintaining SNR greater than 200 by adjusting the exposure time between \SI{0.2}{\milli\second} and \SI{500}{\milli\second}. With our camera capable of exposure times of \SIrange{0.01}{5000}{\milli\second}, we expect that a dynamic range of $100,000:1$ can be readily achieved.

\section{\label{sec:design_analysis}Polarimeter model, error analysis, and calibration}

In this section, we model our polarimeter with Mueller matrices and analyze polarization measurement errors caused by the imperfect robotic movement as well as manufacturing imperfections of optical components being used. We assume that the error of translating the polarimeter impacts negligibly on the polarization measurement, while the imperfect orientation of the polarimeter is modelled by the the tilted waveplates and the polarization axis misalignment.

\subsection{Polarimeter model}

Our imaging polarimeter is modelled by the Mueller matrices of the polarizer $\mathbf{M}_{\text{P}}$ and waveplates $\mathbf{M}_{\text{W}}$ as
\begin{equation}
\mathbf{M}(\theta_{P},\theta_{H},\theta_{Q};\phi_{H},\phi_{Q}) = \mathbf{M}_{\text{P}}(\theta_{P})\mathbf{M}_{\text{W}}(\theta_{H};\phi_{H})\mathbf{M}_{\text{W}}(\theta_{Q};\phi_{Q}).
\end{equation}
The polarizer and waveplates are parametrized by the azimuthal rotation angle $\theta$ and phase retardance $\phi$:
\begin{equation}
\begin{split}
\mathbf{M}_{\text{W}}(\theta;\phi) &= \begin{bmatrix} 1 & 0 & 0 & 0 \\
								0 & C^{2}+S^{2}\cos\phi & CS(1-\cos\phi) & -S\sin\phi \\ 
								0 & CS(1-\cos\phi) & C^{2}\cos\phi+S^{2} & C\sin\phi \\ 
								0 & S\sin\phi & -C\sin\phi & \cos\phi \\ \end{bmatrix},\\	
\mathbf{M}_{\text{P}}(\theta) &= \begin{bmatrix} 1 & C & S & 0 \\
												C & C^{2} & CS & 0 \\ 
												S & CS & S^{2} & 0 \\
												0 & 0 & 0 & 0 \\\end{bmatrix}.
\end{split}
\end{equation}
Here, $C$ and $S$ are $\cos(2\theta)$ and $\sin(2\theta)$, respectively. With $\mathbf{M}_{\text{P}}$ we assume the Wollaston polarizer differs negligibly from perfectly polarizing. Ideally, the phase retardances of the HWP and QWP are $\phi_{H}=\pi$ and $\phi_{Q}=\pi/2$, respectively. We model the polarization extinction between extraordinary and ordinary paths of the Wollaston polarizer by the rotation of the polarizer $\theta_{P}\in\{\SI{0}{\degree},\SI{90}{\degree}\}$. Also, for complete tomography, the rotation angle of the waveplates are in corresponding  pairs of $\theta_{H,Q}\in\{(0\si{\degree},0\si{\degree}), (22.5\si{\degree},45\si{\degree}), (0\si{\degree},45\si{\degree})\}$. The optical power for each combination of the rotation angles can be calculated by multiplying the first row of the Mueller matrix $\vec{M} = [M_{00},M_{01},M_{02},M_{03}]^\top$ to the incident Stokes parameter $\vec{S}_{\text{in}}$. As we have three rotation angle settings of the waveplates and two ports of the polarizer, the six power measurements can be described by the $6\times4$ matrix $\mathbf{A}=[\vec{M}^{(H)},\vec{M}^{(V)},\vec{M}^{(D)},\vec{M}^{(A)},\vec{M}^{(R)},\vec{M}^{(L)}]^\top$ called an \textit{instrument matrix}. Here, the superscript $(i)$ represents each configuration of the polarimeter settings for the power measurements in the horizontal, vertical, diagonal, anti-diagonal, right-circular, and left-circular polarization-basis states. Then, the six power values $\vec{P}$ for the input polarization state can be written as
\begin{equation}\label{eq:intensity}
\vec{P}_{\text{meas}} = \mathbf{A}\cdot\vec{S}_{\text{in}} + P_{d},
\end{equation}
where we added a constant value $P_{d}$ for randomly fluctuating power noises from the camera including the dark current, shot noise, and stray light. Then, the Stokes vector $\vec{S}_{\text{meas}}=\mathbf{W}\cdot\vec{P}_{\text{meas}}$ is obtained from the measured power vector $\vec{P}$ multiplied by the pseudoinverse of the instrument matrix called a \textit{data reduction matrix} $\mathbf{W}=(\mathbf{A}^\top\cdot\mathbf{A})^{-1}\cdot\mathbf{A}^\top$. The obtained Stokes vector is used to reconstruct the density matrix $\hat{\rho}_{\text{out}}$ of the measured polarization state:
\begin{equation}\label{eq:density_matrix}
\hat{\rho}_{\text{out}} = \frac{1}{2}\bigg[\hat{\mathbb{1}} + \frac{S_{1}}{S_{0}}\hat{\sigma_{z}} + \frac{S_{2}}{S_{0}}\hat{\sigma_{x}} + \frac{S_{3}}{S_{0}}\hat{\sigma_{y}}\bigg],
\end{equation}
where $\hat{\mathbb{1}}$ is the $2\times2$ identity matrix.

\subsection{Polarization-measurement error analysis}

Based on the above model, we study how orientation of the robot's end effector ($\alpha$, $\beta$, $\gamma$) changes the reconstructed density matrix $\hat{\rho}_{\text{out}}$. First, it is obvious that the $\gamma$ rotation causes misalignment of the incident polarization state with respect to the principal axes of the waveplates and the polarizer, as depicted in Figure~\ref{fig:polarimeter_description}(a). This can be modelled by equally adding the robot's rotation angle $\gamma$ to the azimuthal angles as the waveplates and polarizer rotate altogether: $\theta_{P(Q,H)} \rightarrow \theta_{P(Q,H)} + \gamma$. The $\alpha$ and $\beta$ rotations are related to the tilt angle of the waveplates $\psi = \cos^{-1}\big(\cos(\alpha)\cos(\beta)\big)$. The phase retardance of the waveplates for a given tilt angle $\psi$ and azimuthal rotation angle $\theta$ is expressed in a closed form~\cite{Gu2018}
\begin{equation}
\begin{split}
\phi(\theta,\psi)&=\frac{2\pi}{\lambda}d\biggl(\sqrt{n_{e}^2-\frac{n_{e}^2\cos^2(\theta)+n_{o}^2\sin^2(\theta)}{n_{o}^2}\sin^2(\psi)}\\&-\sqrt{n_{o}^2-\sin^2(\psi)}\biggr),
\end{split}
\end{equation}
where $\lambda$ is the wavelength of the incident light, $d$ is the thickness of the waveplate, and $n_{o}$ and $n_{e}$ are the ordinary and extraordinary refractive indices, respectively. Here we considered a single-crystal waveplate for simplicity. 

\begin{figure}[t]
	\centering
	\includegraphics[width=\linewidth]{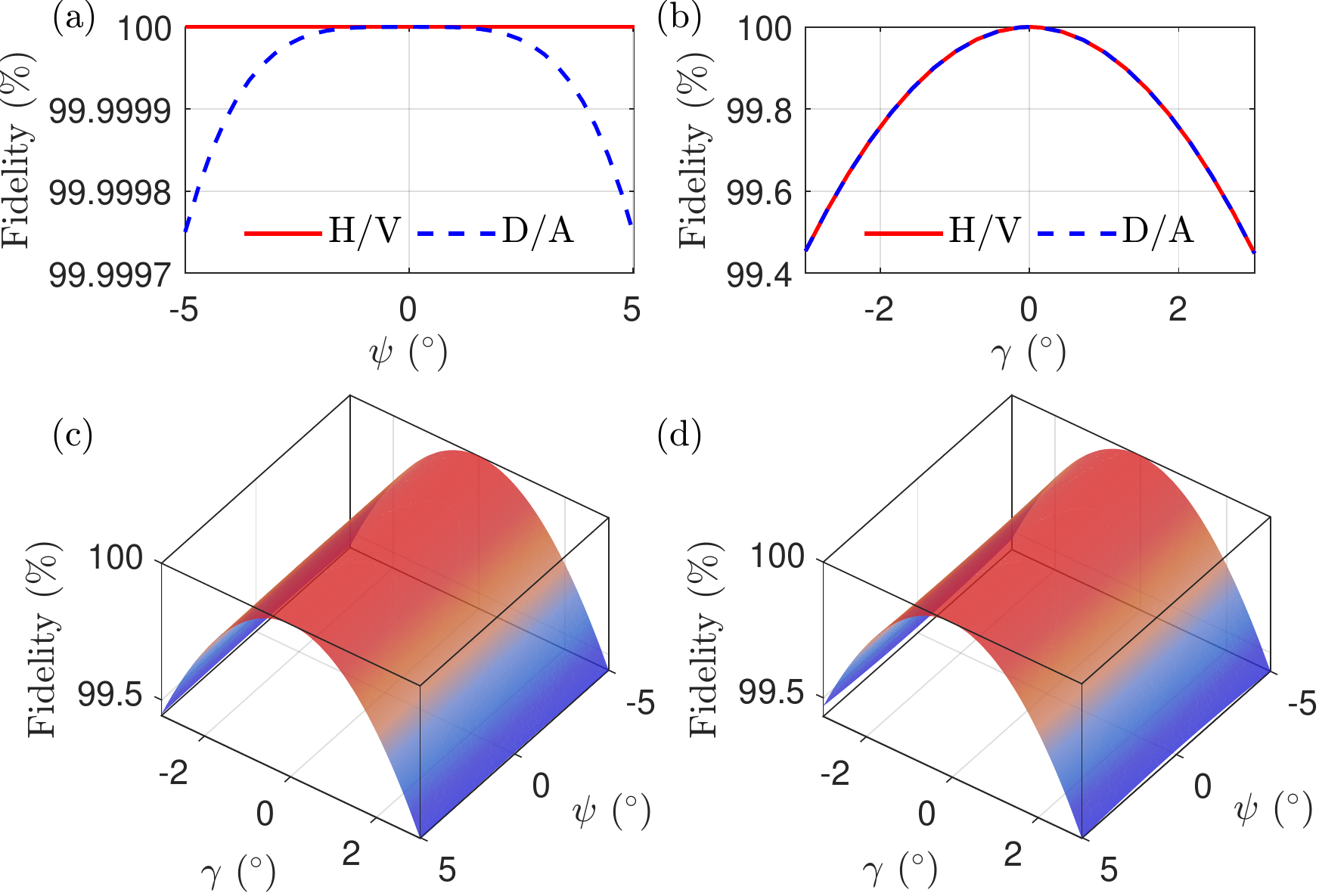}
	\caption{Quantum state fidelity between the measured state by tilted polarimeters and the input state. The fidelity is calculated by our theoretical model for four linearly polarized states as a function of (a) the tilt-angle $\psi$ and (b) the rotation-angle $\gamma$. (c)--(d) three-dimensional plot for the calculated fidelity.}
	\label{fig:motion_polarizationerror}
\end{figure}

We evaluated the quantum state fidelity of the reconstructed density matrix in Eq.~\ref{eq:density_matrix} for the four linear input polarizations as a function of the rotation angle $\gamma$ and the tilt angle $\psi$, as shown in Figure~\ref{fig:motion_polarizationerror}. In our calculation, we modelled the ideal MgF$_{2}$ single-crystal zeroth-order QWP and HWP operating at a wavelength of \SI{785}{\nano\meter}: $d_{H}$ = \SI{33.6}{\micro\meter}, $d_{Q}$ = \SI{16.8}{\micro\meter}, $n_{e}$ = 1.3869, and $n_{o}$ = 1.3752. We found that the fidelity is degraded mainly due to the $\gamma$ rotation and it scales quadratically, whereas the effect for the $\psi$ rotation is relatively negligible.

\subsection{Polarimeter calibration}

\begin{figure}[t]
	\centering
	\includegraphics[width=0.9\linewidth]{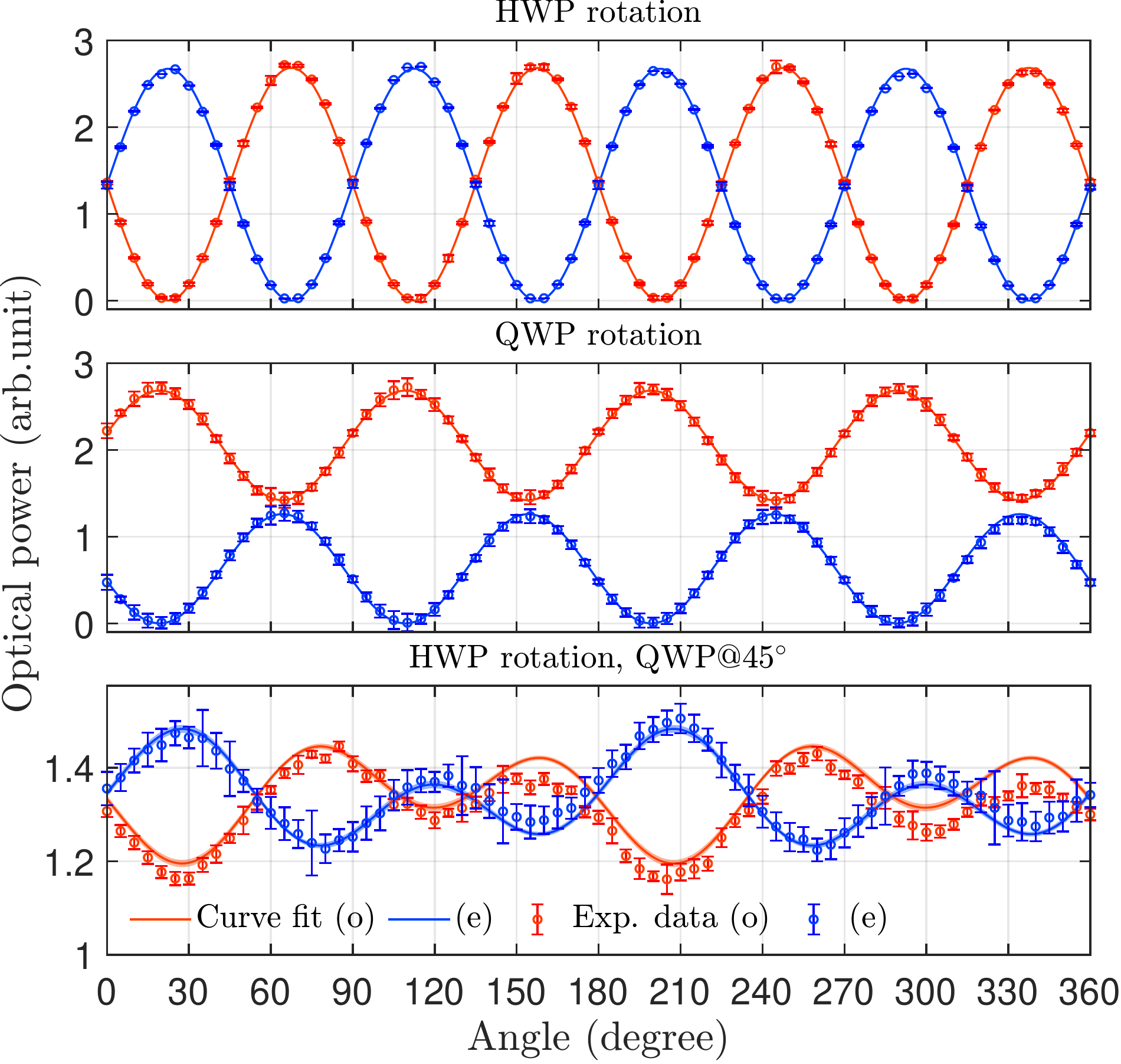}
	\caption{Experimental results of calibrating the polarimeter. Optical powers are measured under the rotation of half- and quarter-wave plates. The incident light is horizontally polarized. The circles show the mean values of twenty power measurements of the ordinary and extraordinary light. The solid curves are fits to the model, and the shaded regions represent the \SI{95}{\percent} confidence interval for the fitting curve. Top: the half-waveplate (HWP) is rotated without the presence of QWP. Middle: the QWP is rotated while the HWP is aligned to the horizontal polarization axis. Bottom: the HWP is rotated while the QWP is oriented at \SI{45}{\degree} with respect to the incident polarization direction.
	}
	\label{fig:calibrate_polarimeter}
\end{figure}

We calibrate our polarimeter by a conventional method~\cite{Boulbry:07} to obtain the instrument matrix including the manufacturing imperfections of the HWP and QWP as well as any other systematic error such as the waveplate misalignment. We aligned the waveplates and the polarizer by using a visible laser field at \SI{532}{\nano\meter} wavelength such that the reflected field is overlapped with the incident field, ensuring that the surface normal vector is parallel to the incidence direction of the laser field. We then injected a horizontally polarized input state at \SI{785}{\nano\meter} wavelength, and recorded the optical powers of the ordinary and extraordinary rays of the polarizer as a function of the rotation angle of the waveplates in three different configurations, as shown in Figure~\ref{fig:calibrate_polarimeter}. First, we rotated the HWP without the QWP to find the angle for the principal axis of the HWP (top). Then, we rotated the QWP while the axis of the HWP has aligned to the incident horizontal polarization (middle). Finally, we rotated the HWP through \SI{360}{\degree} while the optic axis of the QWP was rotated by \SI{45}{\degree} with respect to the incident polarization direction (bottom). Data were collected at \SI{5}{\degree} increments. For each waveplate setting, we captured twenty frames of images to calculate the mean values and standard deviations of the optical power for the ordinary and extraordinary rays of the Wollaston polarizer. Also, we maintained SNR greater than two hundred via auto-exposure time control. We found that the errors in the power measurement are mainly attributed to the background noises in the camera, e.g., dark current and readout noise.

To determine the phase retardance $\phi_{H,Q}$ and misalignment $\theta_{H0,Q0}$ of the waveplates, we used a least-squares fit of the 438 measured power values to our polarimeter model in Eq.~\ref{eq:intensity} with fitting parameters $\{\phi_{H}, \phi_{Q}, \delta\theta, \theta_{H0}, \theta_{Q0}, P_{d}, P_{e}, P_{o}\}$, as shown in Figure~\ref{fig:calibrate_polarimeter}. Here, $\theta_{H0}$ and $\theta_{Q0}$ are the azimuthal angles of the HWP and QWP where their optic axes are aligned to the horizontal polarization. $\delta\theta$ is the azimuthal rotation error between the HWP and QWP due to potential offsets of the two rotation stages. We noticed that the power-measurement efficiencies were slighly different at the two orthogonal basis states; $P_{e(o)}$ quantifies these differential incident optical powers. The difference may be attributed to polarization-dependent quantum efficiency of the camera or imbalanced transmission of the Wollaston polarizer. The fit parameters are $\{\phi_{H}=3.1872\si{\radian}, \phi_{Q}=1.6292\si{\radian}, \delta\theta = -0.0137\si{\radian}, P_{d} = 4.6423\times10^{-6}, P_{0,T}=2.6829, P_{0,R}=2.6741\}$. The \SI{95}{\percent} confidence intervals for $\phi_{H}$, $\phi_{Q}$, and $\delta\theta$ are less than \SI{1.0E-4}{\radian}. The corrected instrument matrix is
\begin{equation}\label{eq:corrected_instrumentMatrix}
\mathbf{A}_{\text{C}} = \begin{bmatrix} 0.5000(0) & 0.5000(0) & 0.0000(1) & 0.0000(3) \\
										     			0.5000(0) & -0.5000(0) & 0.0000(1) & 0.0000(3) \\
				    									0.5000(0) & -0.0032(1) & 0.5006(0) & -0.0124(3) \\
				    						     			0.5000(0) & -0.0032(1) & -0.5006(0) & 0.0124(3) \\
				    									0.5000(0) & 0.0296(0) & 0.0129(3) & -0.4990(0) \\
				    									0.5000(0) & -0.0296(0) & -0.0129(3) & 0.4990(0) \\ \end{bmatrix}.
\end{equation}
To test repeatability, we performed the calibration process five times after repositioning the robot arms from different initial poses: the relative variation of the extracted phase-retardance was measured to be less than \SI{0.2}{\percent}. We repeatedly observed that the theoretical curve is deviated from the measured data, as shown in the bottom plot of the Figure~\ref{fig:calibrate_polarimeter}. Though the discrepancy is small, its origin is presently not understood.

We performed Monte-Carlo analysis to estimate the total polarization measurement uncertainty including both phase-retardance error of the waveplates and motion-induced polarization error, incorporating the measured values of $\phi_H$ and $\phi_Q$. We adjusted the thickness of the waveplates to match the phase retardance to the experimentally obtained values ($\phi_{H}=3.1872\si{\radian}$ and $\phi_{Q}=1.6292\si{\radian}$). We sampled one hundred thousand uniformly distributed random values for the robot arm's orientation error from $\alpha,\beta,\gamma \in$ (-\SI{1}{\degree},+\SI{1}{\degree}) and obtained root-mean-square quantum state fidelity deviation and QBER of \SI{0.01}{\percent} and \SI{0.05}{\percent}, respectively. 

\begin{figure}[t]
	\centering
	\includegraphics[width=\linewidth]{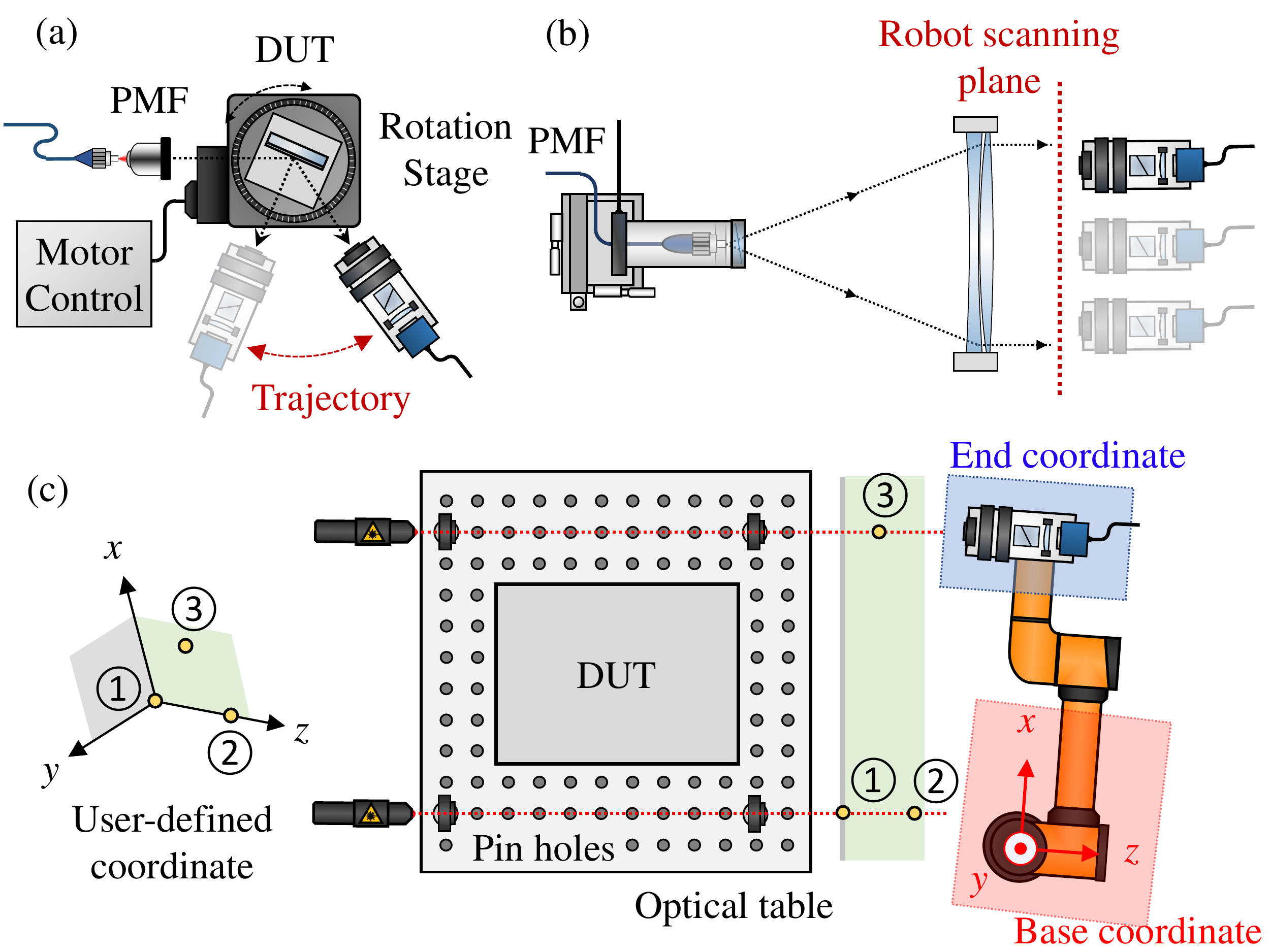}
	\caption{A schematic diagram of the experiment setup for (a) the angle-dependent polarization charaterization of the reflective optics and (b) the characterization of a large lens with the fixed angle of incidence. (c) Three coordinate systems of the robot arm: base, end, and user-defined coordinates. An example method of the user-defined coordinate calibration is illustrated. Three points, one at the origin \raisebox{.5pt}{\textcircled{\raisebox{-.9pt} {\small 1}}}, another on the z-axis \raisebox{.5pt}{\textcircled{\raisebox{-.9pt} {\small 2}}}, and the other on the xz-plane \raisebox{.5pt}{\textcircled{\raisebox{-.9pt} {\small 3}}}, define a user-defined coordinate system. The detailed description of the test procedure is presented in the main text.
	}
	\label{fig:experiment_setup}
\end{figure}

\section{Experimental setup} \label{sec:setup}

We perform a proof-of-principle demonstration of the polarization characterization for a commercial off-the-shelf mirror and a \SI{20.3}{\centi\meter} custom-designed lens. For the mirror, we measure the output polarization states as a function of the reflection angle by rotating the test optic with a fixed incident beam direction. The robot arm moves the polarimeter to track the reflected light from the test optic, and the tomographic measurement is performed at each incident angle, as depicted in Figure~\ref{fig:experiment_setup}(a). For the large lens, we placed an optical fiber at the focal position which produces diverging light with well-defined polarization states. We precisely adjusted the position and angle of the fiber while monitoring the wavefront of the transmitted field by using our aberration characterization system~\cite{CollimationTestPaper}. The incidence direction of the light was aligned to the principal axis of the lens and the divergence angle of the transmitted light was measured to be less than $\pm$\SI{10}{\micro\radian}. We added a \SI{20}{\centi\meter}--diameter mask in front of the lens to block stray light because the numerical aperture of the fiber used in the experiment was larger than the lens. The robot arm moves the polarimeter in the transverse direction (as in a raster scan) to perform the tomographic measurement across the entire output beam, as shown in Figure~\ref{fig:experiment_setup}(b). Then, the quality parameters, e.g., fidelity and purity, were calculated at each position, and the overall quality of polarization maintenance was derived by the median value over the whole aperture.

The robotic manipulator has two pre-set coordinate systems: base coordinates and end coordinates, and the option to set a user-defined coordinate system, as shown in Figure~\ref{fig:experiment_setup}(c). The base and end coordinate is referenced to the absolute position $\{x, y, z\}$ and the orientation $\{\alpha, \beta, \gamma\}$ of the end effector, respectively. These six parameters and the six joint angles can be transformed to each other via forward and inverse kinematics. For testing purposes, it is convenient to define the robot's trajectory in a coordinate whose one axis is parallel to the light propagation direction, which was chosen to be the z-axis. The accurate alignment of this coordinate system is essential to ensure the polarimeter follows the desired path of scanning across the test optic. Following is an example of the procedure to determine the user-defined coordinate.

The three orthonormal bases required to specify a user-defined coordinate system can be determined by three points (assuming a right-handed system), as shown in Figure~\ref{fig:experiment_setup}(c); one point at the origin \raisebox{.5pt}{\textcircled{\raisebox{-.9pt} {\small 1}}}, another point along the z-axis \raisebox{.5pt}{\textcircled{\raisebox{-.9pt} {\small 2}}}, and a final point anywhere on the xz-plane \raisebox{.5pt}{\textcircled{\raisebox{-.9pt} {\small 3}}}. We placed two pairs of two pinholes at the same height on the optical table. The four pinholes define two lines parallel to the surface of the optical table. We shined collimated diode laser light through the pinhole pairs to the polarimeter on the end effector of the robot arm. With one pair of pinholes, we manually centered image spot centroids using the robot teach pendant, setting the aligned position as the origin point. The polarimeter was then moved further away from the table and aligned to the laser light to define the point along the z-axis. Finally, the polarimeter was aligned to the second set of pinholes and the position recorded as a point on the xz-plane. In this way, the orientation of the end effector is aligned such that the polarimeter faces the incident beam, and thus the angles $\alpha$ and $\beta$ are inherently calibrated. The angle $\gamma$ is defined by the incident horizontal polarization axis, and the calibrated polarimeter is oriented such that the power measured at the vertical polarization state is minimized.

\begin{figure}[t]
	\centering
	\includegraphics[width=0.9\linewidth]{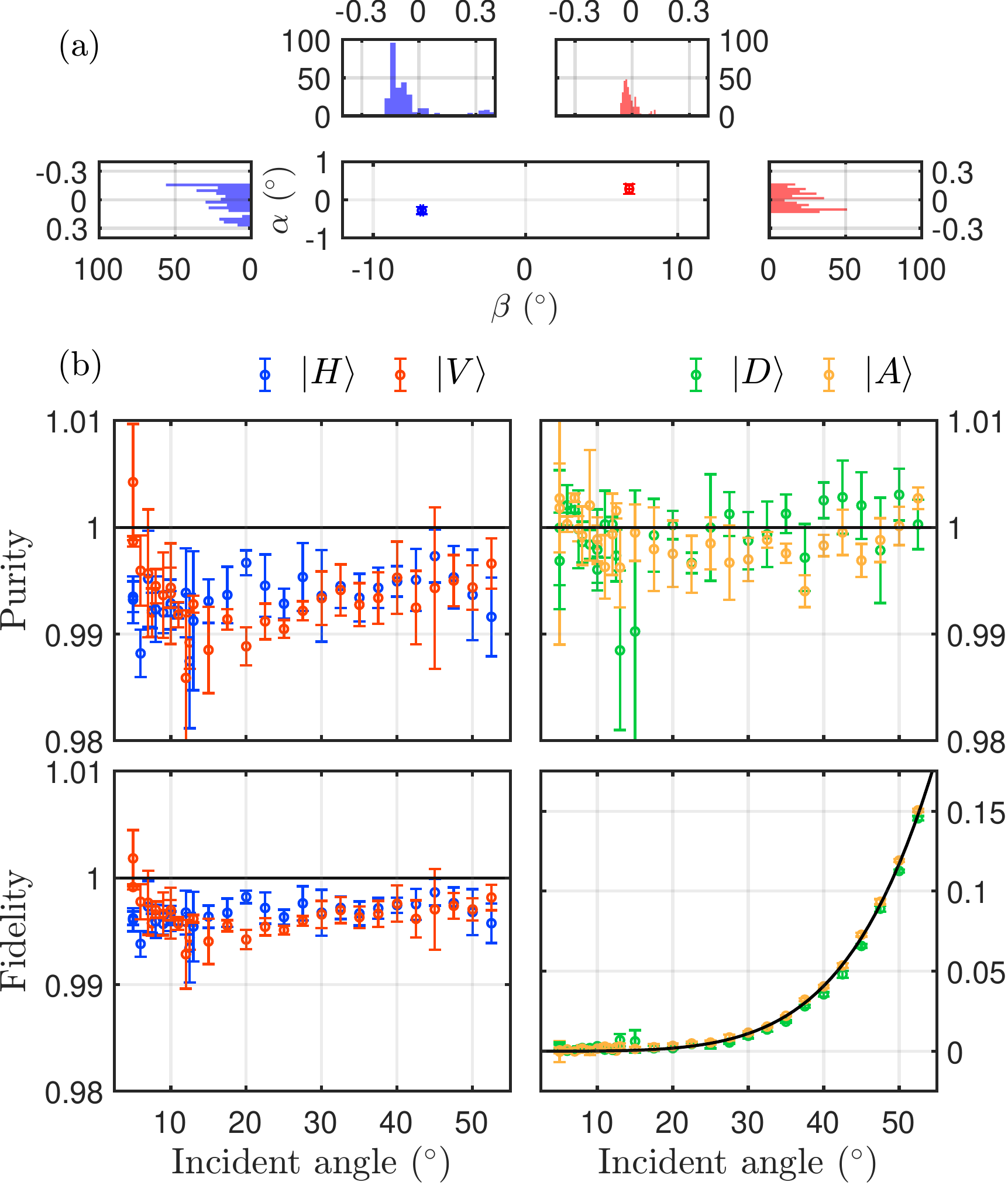}
	\caption{Experimental results from angle-dependent polarization characterization of the protected silver-coated mirror: (a) variation of AOI and (b) fidelities and purity of the measured states for incident horizontal $|H\rangle$, vertical $|V\rangle$, diagonal $|D\rangle$, and anti-diagonal $|A\rangle$ states. Circles show the measured data and solid black lines indicate theoretical predictions based on multilayer thin-film coating calculations.}
	\label{fig:silver_mirror_result}
\end{figure}

\section{Results}\label{sec:results}

\subsection{Commercial off-the-shelf silver-coated mirror} \label{sec:silver-mirror-test}

We measured the polarization states of the reflected light from a \SI{10}{\centi\meter}--diameter protected silver-coated mirror (48-118-557, Edmund Optics) as a function of the reflection angle. Collimated light with \SI{4}{\milli\meter} $1/e^{2}$ beam diameter was sent to the center of the mirror. The horizontal and vertical states of the input polarizations were aligned to p- and s-polarizations of the mirror, respectively. We rotated the mirror to vary the reflection angle from \SIrange{5}{52.5}{\degree}, and the robot arm follows the reflected beam in an arced path. The trajectory was defined by the measured radius from the center of the mirror to the center of the end effector with a laser-distance measurer. At each angle, twenty frames were captured to calculate the mean and standard deviation of the measured powers and centroids while maintaining SNR above one hundred. The full scan of the polarization test was repeated four times for statistical certainty.

Figure~\ref{fig:silver_mirror_result}(a) shows the variation of the AOI to the polarimeter during the test. AOI variation was maintained within $\pm$\SI{0.2}{\degree}, indicating good coordinate alignment and excellent repeatability. As shown in Figure~\ref{fig:silver_mirror_result}(b), the averaged purity and fidelity for horizontal and vertical input polarization states are maintained above \SI{99}{\percent} over the entire reflection angle range, indicating good alignment of horizontal and vertical polarization states to the s- and p-polarizations of the mirror.

The polarization-dependent reflectivity of the protected silver mirror can be accurately calculated by multilayer coating calculations~\cite{WEINSTEIN19543}. It is expected that the comparison between experimental data and theoretical prediction validates our methodology, i.e., the usage of robot arm to move the polarimeter for angle-dependent polarization characterization. We modelled the protected silver film with a \SI{92}{\nano\meter}--thick SiO$_{2}$ layer coated on top of a \SI{1}{\micro\meter}-thick silver layer. The transmissive and reflective coefficients of the s- and p-polarizations were calculated by a conventional optical admittance method~\cite{10012538968}. The coefficients were then used to obtain the polarization states of reflected light. In Figure~\ref{fig:silver_mirror_result}(b), black lines indicate theoretical values of fidelity and purity as a function of reflection angle. We characterize a closeness between theory and experiment by the standard error of regression $\text{SER}=\sqrt{\sum_{i=1}^{n}(y_{i}-f(x_{i}))^2/\left(n-k\right)}$ with $k$ denoting the number of free parameters for the theoretical model. In our case, we consider the thickness of SiO$_{2}$ and silver layer as free parameters, and thus $k=2$. Here, $y_{i}$ and $f(x_{i})$ are the measured and theoretically prediected values, respectively. We calculated SER of fidelity being better than \SI{0.4}{\percent} for all four input polarization states, showing the excellent agreement between theory and experiment. 

In our experiment, the density matrices of measured polarization states are reconstructed by calculating the Stokes vector that is normalized by the total intensity averaged over three different polarization measurement bases. As we used collimated light with beam diameter of \SI{4}{\milli\meter} and set the pin-hole size to be \SI{2}{\milli\meter}, any positional instability while rotating the waveplates causes total intensity variation. This effect may yield unphysical quantum states whose purity is greater than unity, as seen in Figure~\ref{fig:silver_mirror_result}(b). This issue can be resolved by either larger pinhole size or other alternative tomographic reconstructions such as the maximum likelihood estimation.

\begin{figure}[t]
	\centering
	\includegraphics[width=\linewidth]{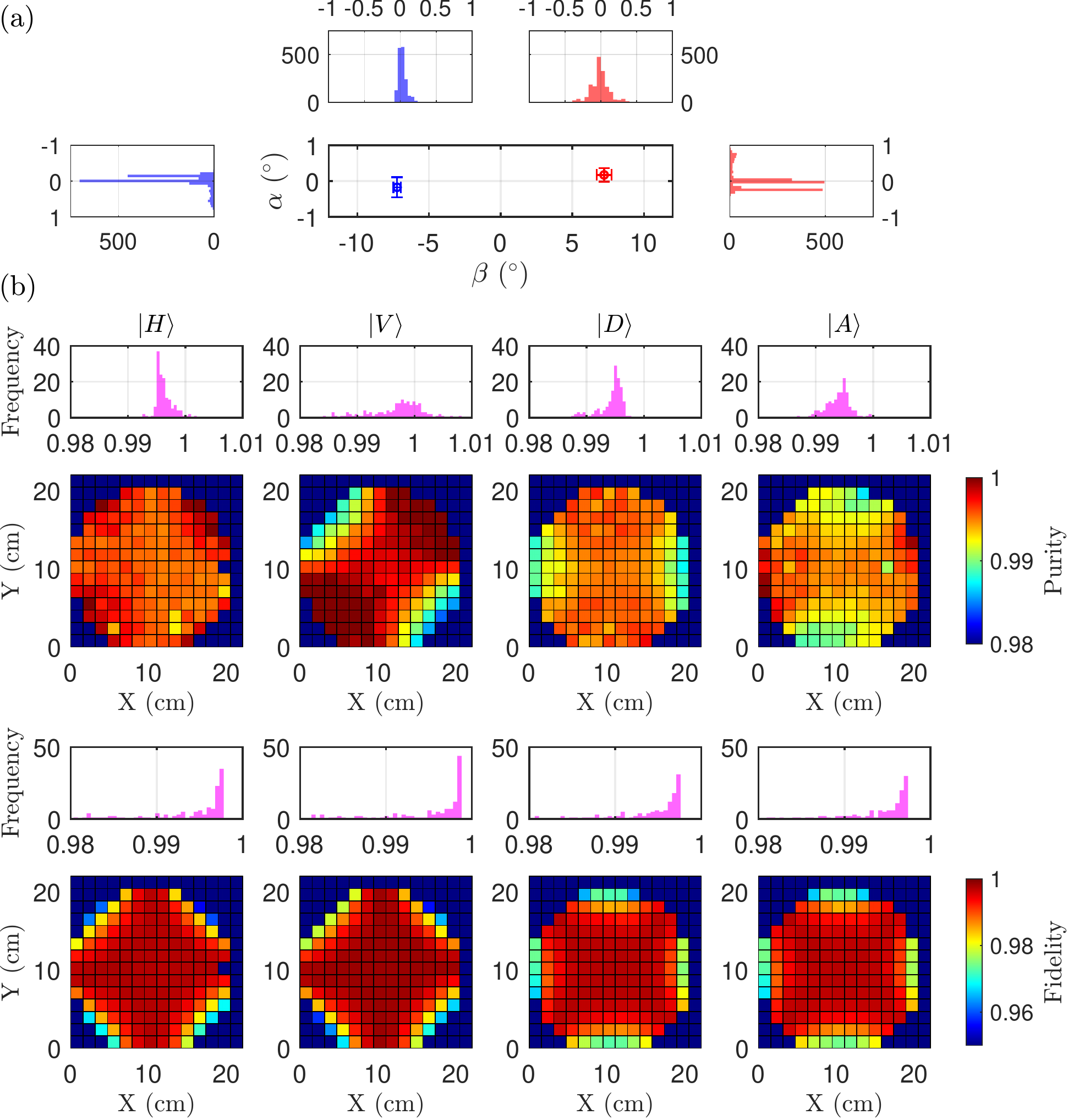}
	\caption{Experimental results from polarization test of the large convex lens: (a) variation of AOI and (b) color maps of fidelity and purity of the measured states for incident horizontal $|H\rangle$, vertical $|V\rangle$, diagonal $|D\rangle$, and anti-diagonal $|A\rangle$ states.
	}
	\label{fig:lens_test_result}
\end{figure}

\subsection{Large custom-designed lens} \label{sec:large-optics-test}

We designed a \SI{20.3}{\centi\meter} cemented doublet with \SI{243.8}{\centi\meter} effective focal length. The lens was manufactured by Hyperion Optics with H-LAF2 and H-ZF5 materials. The size of the lens and the design specifications were determined by quantum link-budget analysis and the impact of the optical aberrations as well as the atmospheric turbulence. Each surface was coated with six layers of Ta$_{2}$O$_{5}$ and SiO$_{2}$ to keep reflectivity below \SI{0.5}{\percent} at \SI{780}{\nano\meter}, \SI{980}{\nano\meter}, and \SI{1550}{\nano\meter} wavelengths. 

In our polarization characterization of the doublet, we pixelized the transverse mode profile of the transmitted light with an even $15\times15$ grid. The robot arm scans through all 225 positions at a given input polarization, and the 2-dimensional scan was repeated for all four different input polarization states. During the scan, the polarization measurement is skipped at the position where the incident light is too weak to be detected with the exposure time greater than \SI{100}{\milli\second}. The sampling size of light during this scan is limited by the iris aperture of \SI{3}{mm}. Since all power measurements at six orthogonal polarization bases are performed at one position and then the robot moves the polarimeter to the next position, the median and quartiles of the quality parameters, i.e., fidelity, purity, and QBER, over the measured area includes the position and orientation error of the robot arm.

\begin{table}[t!]
\caption{\label{table:lens_test_results}Median values of the fidelity, purity, and QBER for the transmitted polarization states from the \SI{20.3}{\centi\meter} custom-designed len. Lower and upper quartiles are listed in parantheses.}
\centering
\begin{ruledtabular}
\begin{tabular}{ccccc}
Input state &	Fidelity (\%)		&	Purity (\%) &	QBER (\%)		 \\
\hline
\multirow{ 2}{*}{$\ket{H}$} 	& 99.60		&	99.59		&	0.40&	\\
& (99.01,99.73)		&	(99.53,99.70)		&	(0.27,0.99)	&	\\
\multirow{ 2}{*}{$\ket{V}$} 	  	& 99.68 		&	99.79		&	0.32\\
& (99.03,99.84) 		&	(99.44,100.00) 		&	(0.16,0.97) \\
\multirow{ 2}{*}{$\ket{D}$} 	  & 99.55		&	99.50		&	0.35 	\\
& (98.99,99.70)		&	(99.38,99.56)	&	(0.19,0.97)	\\
\multirow{ 2}{*}{$\ket{A}$} 	 	& 99.52		&	99.42	&	0.42 \\
& (98.97,99.68)	&	(99.27,99.51)	&	(0.28,1.01) \\
\end{tabular}
\end{ruledtabular}
\end{table}

Figure~\ref{fig:lens_test_result}(a) shows the variation of measured AOI during the polarization test. It was observed that the AOI to the polarimeter was maintained within $\pm$\SI{0.5}{\degree}. The measured purity and fidelity are presented in the color maps shown in Figure~\ref{fig:lens_test_result}(b). Histograms of the measured values indicate the uniformity of transmitted polarization states across the aperture. QBERs for the four incident polarization states were directly calculated from the raw power measurements. The median and quartiles of the three quality parameters, i.e., fidelity, purity, and QBER, are listed in Table~\ref{table:lens_test_results}. The outcome of this test shows great polarization maintenance as the typical fidelities for all four input states are greater than \SI{99.5}{\percent}. The acceptable QBER for ground-to-satellite QKD links\cite{Compre_design_JP} is order of \SI{1}{\percent}, and our promising result shows that the lens is suitable for free-space QKD experiments.

It is worth noting that the high-fidelity region shows a ``cross-mark" feature on the color maps. This feature seemingly depends on the input polarization states; plus--sign in horizontal and vertical state and X--shape in the diagonal and anti-diagonal input states. In our setup, the input polarization states are defined by the rotation of the polarizer and the fiber together, and the high fidelity region is correlated to this rotation angle, indicating that the high- and low-fidelity region may not be attributed to the quality of the test optic, but rather by the uniformity of the input polarization state across the lens aperture. The reason for the imperfect state preparation with the polarized diverging beam will be further investigated. The full characterization for the instrumental polarization of the lens can be performed by directly characterizing the input states with the same polarimeter and comparing the results as in Mueller-matrix polarimetries~\cite{Azzam:16}.

The precision of our imaging polarimeter is mainly limited by the dynamic range and noise of the camera being used. This is indicated by the interquartile range of the measured QBERs that are greater than the median values. Also, note that the linearity of the camera's exposure time showed uncertainty from \SIrange{0.1}{1.6}{\percent}. Since our AOI measurements verified the reliable control of the position and orientation of the polarimeter, the replacement of the camera with two photomultiplier-tube (PMT) modules may be considered in future to improve polarization measurement precision~\cite{Bailey2015}. Indeed, we replaced our imaging polarimeter with a conventional division-of-amplitude polarimeter consisting of the HWP, QWP, polarized beam splitter, and two balanced power meters for the polarization characterization of a prototype telescope for the QEYSSat payload~\cite{Podmore2021}. Across the four input polarization states, the measured QBER in that test was less than \SI{0.05}{\percent}. The QBER can be translated to the polarization extinction ratio greater than \SI{33}{\decibel} which is comparable with the polarization-test results of the telescope~\cite{Wu2017} and optical terminal for the Micius satellite~\cite{Han2020,Wu2020}.

\section{Conclusion}\label{sec:conclusions}

We developed a robotized polarization characterization platform for optical devices in free-space quantum communications. Our system can easily be adjusted for performing polarization tests on diverse reflective or refractive optical systems with a wide range of aperture sizes (up to 30 cm), and either curved or flat surfaces at consistently high precision. The measurement apparatus can be readily set up in outdoor and used for deployed systems. The characterization process is fully automated once the robot's coordinate system is calibrated. Our imaging polarimeter is capable of monitoring the variation of incident angle, and the tilt error of the polarimeter due to the robot's motion can be detected. This feature could be used for implementing a feedback mechanism to correct the polarimeter's position and orientation in the future. We presented our theoretical analysis of the polarization measurement error caused by the tilt and rotation of the polarimeter, and showed that the misalignment of the polarization axis due to the azimuthal rotation is a dominant measurement error. This rotation error could be detected and compensated by injecting more incident states to fully characterize the change of the polarizations, as in conventional Mueller-matrix polarimetries. It is worth noting that the input polarization states can be directly characterized in our system, and therefore the precision of the polarization characterization of the test optic is limited by the measurement device.

We performed a proof-of-principle experiment for the polarization characterization of two different optical components. First, a commercial silver-coated mirror was characterized as a function of the reflection angle. The polarization states of the reflected light were theoretically calculated, and the excellent agreement between the theory and the test results validated our method of moving polarimeter in a pre-determined path with an industrial robotic manipulator. Secondly, the polarization characterization of our custom-designed lens for the QEYSSat mission showed a good polarization preservation across the full aperture. The quality parameters provided a lower bound of the performance of the test optic which includes imperfection of input state preparation. The typical QBER was estimated to be \SI{0.37}{\percent} and the ﬁdelity was greater than \SI{99.5}{\percent}. In both experiments, the angle of incidence to the polarimeter remained within $\pm$\SI{0.5}{\degree}, which showed the reliable control of the polarimeter's position and orientation using the robot arm. 

Our polarization characterization clearly demonstrates the viability of using an industrial robotic manipulator to study large optical components and systems. Our theoretical model and experimental demonstration showed that the motional precision achieved is sufficient to draw robust conclusions from the optical measurements. We believe that our robotized polarization characterization platform could therefore also support the development of free-space optical components or terminals for a broad range of applications including laser communications, lidars, and astronomical observatories. Furthermore, recent developments of polarization imaging cameras could be directly used in our robotized polarization measurement scheme for automation of characterizing other physical properties such as stress measurements and bio-imaging applications~\cite{PolnImageCam}.

\section*{Acknowledgement}
Y.S.L acknowledges support from the Mike and Ophelia Lazaridis Fellowship Program. This research was supported in part by the Canadian Space Agency; Canada Foundation for Innovation (25403, 30833); Ontario Research Foundation (098, RE08-051); Canadian Institute for Advanced Research; Natural Sciences and Engineering Research Council of Canada (RGPIN-386329-2010); Industry Canada.

\section*{Author Declarations}
\subsection{Conflict of interest}
The authors have no conflicts to disclose.

\section*{Data availability}
The data that support the findings of this study are available from the corresponding author upon reasonable request.

\bibliography{Bibliography}

\end{document}